\documentclass[conference]{IEEEtran}
\usepackage{cite}
\usepackage[pdftex]{graphicx}
\usepackage{amsmath}
\usepackage{algorithmic}
\usepackage{array}
\usepackage{caption} 
\usepackage{amsmath}
\newtheorem{theorem}{Theorem}

\newtheorem{definition}{Definition}

\begin{document}
\IEEEoverridecommandlockouts

\title{Time-universal data compression and prediction}
\author{\IEEEauthorblockN{Boris Ryabko}
\IEEEauthorblockA{Institute of Computational Technologies of SB RAS\\ 
Novosibirsk state university
\\Novosibirsk, Russian Federation \\Email: boris@ryabko.net\\}
}

\maketitle

\begin{abstract}
Suppose there is a large file which should be transmitted (or stored) and there are several (say, $m$) 
admissible data-compressors. It seems natural to try all the compressors and then choose the best,
i.e. the one that gives the shortest compressed file. Then transfer (or store)
the index number of the best compressor (it requires $\lceil \log m \rceil $ bits) and the compressed file.
The only problem is the time, which essentially increases due to the need to compress the 
file $m$ times (in order to find the best compressor). We  propose a method that encodes the file with the optimal compressor, but uses a relatively small additional time: the ratio of this extra time and the total time of calculation can be limited by an arbitrary positive constant. 

Generally speaking, in many situations  it may be necessary find the best data compressor out of a given set, which is often done by comparing them empirically.
 One of the goals of this work is to turn such a selection process into a part of the data compression method, automating and optimizing it.

A similar result is obtained for the related problem of time-series forecasting.
\end{abstract}


%
\IEEEpeerreviewmaketitle

\section{Introduction and preliminaries}

\subsection{General description of the problems and results}
Nowadays there are many efficient lossless data-compressors (or archivers)  which are widely used in information technologies.
These compressors  are based on different ideas and approaches, among which we note the PPM universal code \cite{cleary1984data} 
(which is used along with the arithmetic code \cite{rissanen1979arithmetic}), the Lempel-Ziv (LZ) compression methods
\cite{ziv1977universal}, the Burrows-Wheeler transform \cite{burrows1994block} (which is used along with the book-stack (or MTF) code \cite{ryabko1980data,bentley1986locally,ryabko1987technical}) and 
the  class of grammar-based codes \cite{kieffer2000grammar,yang2000efficient}. 
All these codes are universal. This means that, asymptotically, the length of the compressed file goes to the  smallest possible value (i.e. the 
Shannon entropy per letter), if the compressed sequence is generated by a stationary  source.
 
Currently, several dozens of archivers are known, 
 each of which has certain merits and it is impossible to single out one of the best or even remove the worst ones. The main part of them are universal codes  
(as far as a computer program can meet asymptotic properties).  
Thus,  the  one faces the problem of choosing the best method to  compress  a given file. 

Suppose  someone wants to compress a certain  file in order to store it (or transfer it). 
It seems natural to use for compression  the best compressor: the one which gives the shortest compressed file.
 In such a case one can  try to compress the file  in turn by all the compressors 
and then 
store the name of the best compressor (as a prefix) and the file, compressed by the best method. 
 An obvious drawback of this approach is the need to
spend a lot of time in order to 
 first compress the file by all the compressors.
 
 In this paper we show  that there exists a method that encodes the file with the optimal compressor, but uses a relatively small additional time. 
Very briefly, the main idea of the suggested approach is as follows:  in order to find the best,
try all the archivers, 
 but, when doing it,  use for compression only a small part of the file. 
Then apply the best archiver for the compression of the whole file. 
It turns out, that 
under certain conditions on the source of the files, the total time can be made as close to the 
minimal as required. Thus, we call such methods "time-universal". 
This scheme can be extended to 
the problem of  time-series
forecasting,  which  is considered in a framework of the Laplace approach (This approach is  shortly described  in   Appendix 2.)
 
 In this paper we suggest time-universal methods for
 data compression  and forecasting. 
 To the best of our knowledge, the suggested  approach to prediction and compression is new, but close ideas have been considered 
 in algorithmic information theory and  artificial intelligence, where they were developed  for solving other problems \cite{Vitanyi:08,Hutter:04uaibook}.

  \subsection{ The  over-fitting  problem }
  If someone wants to find the best method of prediction or data compression, she/he 
 should take into account the so-called  over-fitting problem. 
The over-fitting problem is the phenomenon in which the accuracy of the model on unseen
data is poor whereas the training accuracy is nearly perfect.


In our case, 
there is a set of either 
data compressors $F = \{ \varphi_1, \varphi_2, ... \} $ or 
predictors $\Pi = \{ \pi_1, \pi_2, ... \} $.
  Besides,  there is a sequence 
$x_1 x_2 ... x_n, n> 1$, and one should choose a good method from the set of predictors (or
data compressors) based on  investigating of  a short  initial part $x_1 x_2 ... x_l, \,  l < n.$
In the case of data compression,  it is natural to choose 
such a method $\hat{\varphi} \in F$, for which $| \hat{\varphi}( x_1 x_2 ... x_l)|$ is minimal. 
In the case of forecasting, it is natural to choose such a predictor $\hat{\pi } \in \Pi$
for which the probability 
 $\hat{\pi}( x_1 x_2 ... x_l)$  is maximal (the maximum likelihood principle.)

In this  situation the problem of over-fitting is as follows: if $x_1 ... x_l$ is a relatively 
short sequence and the set of methods $F$ is large or even infinite, it is possible 
that a performance of the chosen method $\varphi$ on    
 $x_1 ... x_l$ is good, but on
the whole sequence
$x_1 ... x_n$  it is bad.  The over-fitting problem for prediction is similar:
the error  of the chosen predictor on unseen
data is large whereas the training error is nearly zero.  

We consider a solution to this problem based on the approach developed in the theory
of universal coding \cite{BRyabko:84,BRyabko:88}, but note that  a similar solution can be obtained 
in the framework of MDL (minimal description length) method suggested by J.~Rissanen 
\cite{Rissanen:78,Rissanen:89} and developed in numerous papers \cite{barron1998mdl,Grunwald:07,kontkanen2000predictive}.
For this we need such a probability distribution $\omega$ on the set $1, 2, 3, ... $
for which all $\omega_i >0$. 
For example, the following:
   \begin{equation}\label{ome}
 \omega_k = \frac{1}{k(k+1)} \, , \, \,  k = 1, 2, 3, ... \, \, . 
\end{equation}
(Clearly, this is a  probability distribution, because $\omega_k = 1/k - 1 /(k+1) $.)
The described approach to problem of over-fitting  is to find a
data-compressor $\varphi_s$ for which 
$- \log \omega_s \,+ |\varphi_s(x_1 x_2 ... x_l)|$ is minimal:
$- \log \omega_s \, + |\varphi_s(x_1 x_2 ... x_l)|$ $= \min_{i=1,2, ... } $ 
$ (- \log \omega_i \,+| \varphi_i(x_1 x_2 ... x_l)| )$. 
Note that, if the  set of data-compressors is finite,  it is possible to use an uniform distribution
$\omega_i = 1/ |F|$, $i=1, ..., |F|$.
It is worth noting that there is a natural interpretation of the considered solution. The value $ \lceil- \log \omega_i \rceil + |\varphi_i (x_1 x_2 ... x_n)|$
can be considered as a codeword length, where the first part $ \lceil- \log \omega_i \rceil$ encodes the number
$i$, whereas the second part $\varphi_i (x_1 x_2 ... x_n)$ encodes $x_1 x_2 ... x_n$
by the data compressor $\varphi_i$.

In the case of prediction the solution of    the over-fitting problem is similar: 
find a predictor $\pi_s$ for which $\omega_s \pi_s(x_1 x_2 ... x_l)$ is maximal.

\section{  Description of problems and the main notations}
In  this section we first  consider the following problem:
There is a  set of data compressors $F = \{\varphi_1, \varphi_2, ... \}$ and let
$x_1 x_2 ... $ be a sequence of letters from a finite alphabet $A$
whose  
initial part $x_1 ... x_n$ should be compressed by some $\varphi \in F$. 
Let, as before,  $v_i$ be the time spent on encoding 
one letter by the data compressor $\varphi_i$  
 and suppose that all $v_i$ are upper-bounded by a certain constant $v$, i.e. 
$
\sup_{i=1, 2,  ... , } v_i \,  \le  \, v\,   .
$
(Note, that $v_i$ can be unknown beforehand, but $v$ should be known.) 
   
   The goal is to find a good data compressor  from $F$   in order to compress  $x_1 ... x_n$
in such a way that the total time spent for all calculations and compressions
does not 
exceed $T (1 + \delta), \delta >0 $, where $T = v \,n$ is  the
minimum time that must be reserved for compression
 and $\delta T$ is an additional time that can be used to find  the good compressor (among $\varphi_1, \varphi_2, ...$).  In order to accurately describe the problem,  we
suppose also 
  that there is a probability distribution $\omega = $  $\omega_1, \omega_2, ...$ such that all
  $\omega_i >0$.
 The goal is  to fined such $\varphi_i$  that the value 
$$\lceil  - \log \omega_i  \rceil  \, + |\varphi_i (x_1 x_2 ... x_n) | 
$$ is close to minimal.  (Here the first part $\lceil - \log \omega_i \rceil$ is used for encoding number $i$.)
The decoder first finds $i$ and then $x_1 x_2 ... x_n$ using  the decoder 
corresponding $\varphi_i$.
\begin{definition}
We call any method that encodes a sequence $x_1 x_2 ... x_n$, $n \ge 1$, 
$x_i \in A$, by the binary word of the length $ \lceil - \log \omega_j \rceil\, +\, |\varphi_j (x_1 x_2 ... x_n)|$
for some $\varphi_j \in F$, a time-adaptive code and denote it by $\hat{\Phi}^\delta_{compr}$.
The output of $\hat{\Phi}^\delta_{compr}$ is the following word:
 \begin{equation}\label{outp-compr}
 \hat{\Phi}^\delta_{compr}(x_1 x_2 ... x_n)     \,  =    < \omega_i >   \,  \, \varphi_i (x_1 x_2 ... x_n)   \,  ,
\end{equation}
where $< \omega_i >$ is  $ \lceil- \log \omega_i \rceil$-bit word that encodes $i$,
whereas the time of encoding is not grater than $T(1+\delta)$.

If for a time-adaptive code $\hat{\Phi}^\delta_{compr}$ the following equation is valid
$$
 \lim_{t \to \infty}  \hat{\Phi}^\delta_{compr}(x_1 ... x_t)  / t \,= \inf_{1=1, 2,  ... } 
  \lim_{t \to \infty}  \varphi_i(x_1 ... x_t)  / t \, , 
$$
this code is called time-universal.

The definition   
for the forecast is as follows:
Let there be   a  set of predictors $\Pi = \{\pi_1, \pi_2, ... \}$. 
 By definition, the goal of the time-adaptive predictor
 $ \hat{\Phi}^\delta_{pred} $ is  to spend the extra time $\delta \, T$ in order  to find such $\pi_i$  that the value 
$$  \omega_i   \,  \pi_i (x_1 x_2 ... x_n) 
$$ is close to maximal.
By definition, 
 the output of the time-adaptive predictor  $\hat{\Phi}^\delta_{pred}$ is the following set of forecasts (conditional probabilities):
 $$   
 \{ \pi_j (a|x_1 ... x_n), \, \,  a \in A \},
$$
for a certain $\pi_j \in \Pi$.
It will be 
convenient  to define
\begin{equation}\label{outp-pr}
 \hat{\Phi}^\delta_{pred}(x_1 x_2 ... x_n)     \,  =  \,   \omega_i \, 
  \pi_j (x_1 ... x_n) \, \, .
\end{equation}
If for a predictor $\hat{\Phi}^\delta_{pred}$ the following equation is valid
$$
 \lim_{t \to \infty} (- \log \hat{\Phi}^\delta_{pred}(x_1 ... x_t)  )/ t = 
 $$ $$\inf_{1=1, 2,  ... } 
  \lim_{t \to \infty} ( - \log  \pi_i(x_1 ... x_t)  )/ t \, , 
$$ and, for any $t$, time of calculation is not grater than $T(1+ \delta ) $
this predictor is called time-universal.

\end{definition}

\textbf{Comment 1.} 
 Here and below we did not take into account the time required for the calculation of  
$\log \omega_i$ and some other 
auxiliary calculations. If in a certain situation this time is not negligible,  
it is possible to reduce $\hat{T}$ in advance by the required value.

\section{Finite number of data-compressors or predictors}
Suppose that there is a file $x_1 x_2 ... x_n$ and data compressors $\varphi_1, ..., \varphi_m$, 
$n \ge 1, m \ge 1$. 
Let, as before,  $v_i$ be the time spent on encoding 
one letter by the data compressor $\varphi_i$, 
 \begin{equation}\label{tau}
v = \max_{i=1, ... , n} v_i, \, \, \,   T = n  \, v\,   ,
\end{equation}
and let 
 \begin{equation}\label{T}
 \hat{T}= T (1+\delta) \, , \, \, \delta > 0.
\end{equation}
The  goal is to find the data compressor $\varphi_j$, $j=1, ... , m$,
  that compresses the file
$x_1 x_2 ... x_n$  in the best way in time 
$\hat{T}$.
Seemingly, the simplest method is as follows: 

\textbf{Step 1} Calculate $r = \lfloor \delta T / v \rfloor $.

\textbf{Step 2}  Compress the file $x_1 x_2 ... x_r$ by $\varphi_1$ and find the length of compressed file
$| \varphi_1 (x_1 ... x_r)|$,  then , likewise, find $| \varphi_2 (x_1 ... x_r)|$, 
$| \varphi_3 (x_1 ... x_r)|$, etc.  

\textbf{Step 3} Calculate $ s = \arg \min_{i=1, ... , m} $ $| \varphi_i (x_1 ... x_r)|$ 

\textbf{Step 4} Compress the whole file $x_1 x_2 ... x_n$
by $ \varphi_s $ and compose the codeword $ \langle s \rangle$  $\varphi_s(x_1 ... x_n)$,
where $\langle s \rangle$ is $\lceil \log m \rceil  $-bit word with the presentation  of $s$.

The decoding is obvious.  Denote this method by $\Phi^\delta_1$.

\textbf{Comment 2.}
We considered the case of data compression. It is possible to apply the described method
for time-universal prediction.  In this case one should calculate $\pi_i(x_1 ... x_r)$
instead of $| \varphi_i (x_1 ... x_r)|$ and the third step should be changed as follows:

Calculate $ s = \arg \max_{i=1, ... , m} $ $ \pi_i (x_1 ... x_r).$

The asymptotic properties of the method $\Phi_1^\delta$ are as follows:

\textbf{Claim 1.}  Let there be an infinite sequence $x_1, x_2, ... $ and data 
compressors $\varphi_1, ... , \varphi_m$ 
such that there exist the following limits 
\begin{equation}\label{lim}
\lim_{n \to \infty} |\varphi_i(x_1 x_2 ... x_n)| / n   \, \, , 
\end{equation}
for all $i = 1, ..., m$. 
Then, for any $\delta > 0$
$$
\lim_{n \to \infty} |\Phi^\delta_1(x_1 x_2 ... x_n)| / n   \, \,  = 
\min_{1, ... , m} 
\lim_{n \to \infty} |\varphi_i(x_1 x_2 ... x_n)| / n   \, \, ,
$$
i.e.  $\Phi^\delta_1$ is time-universal.

Next we describe a more general method, for which Claim 1 is a special case.

\section{General method}
Generally speaking, it is possible to offer many reasonable strategies for finding the optimal data compressor (or predictor) for a given time.
For the finite set of data-compressors  such a strategy  can be as follows:
try all the compressors on a (very) short sequence $x_1 x_2 ... x_k$ 
 and choose a few of the best ones.
Then try those chosen data-compressors
 on  a larger sequence $x_1 x_2 ... x_l$, $k < l$, and choose the best
which will be used for compression of the whole sequence $x_1 x_2 ... x_n$. 
Another reasonable strategy can be based on maximization of the probability to determine
the optimal data compressor as a function of the extra time $\delta \, T$ and other parameters.

So, we can see that there are a lot of reasonable strategies and each of them has a lot of parameters.
That is why, it could be useful to use 
multidimensional optimization methods, such as  machine learning, 
 so-called deep learning, etc.  
Since this is the first paper devoted to time-adaptive  
and time-universal
data compression and prediction, we consider only some general conditions
needed for time-universality.

For a time-adaptive data-compressor 
$\hat{\Phi}$ and $x_1 ... x_t$ 
we 
define  for any $\varphi_i$  
$$
\tau_{\varphi_i} (t) = \max  \{r:  \varphi_i (x_1 ... x_r)  \, \, is \, \, caculated , 
$$ $$ \, \, when
\, \,  \hat{\Phi}(x_1 ... x_n) \, \, is \, \, applied.
$$
\begin{theorem}
If the following properties are valid:

i)    for all $i = 1, 2, ... $
 \begin{equation}\label{th-i1}
 \lim_{t \to \infty} \tau_i(t) = \infty ,
 \end{equation}
ii)  for any $t$ the method $\hat{\Phi}$ uses such a compressor $\varphi_{s(t)}$ 
for which,   for any $i$ and $r = \min \{\tau_i, \tau_{s(t)} \}$
   \begin{equation}\label{th-i2}
- \log \omega_{s(t)} + |\varphi_{s(t)} (x_1 ... x_{r}) |  \le 
- \log \omega_{i} + |\varphi_{i} (x_1 ... x_{r}) |\, ,
 \end{equation}
iii)  the  limits  $\lim_{t \to \infty}  \varphi_i(x_1 ... x_t)  / t $
exist for all $\varphi_i$.

Then  $\hat{\Phi}(x_1 ... x_n) $ is time universal, i.e.,
 in a case of data compression, 
 \begin{equation}\label{th-claim}
 \lim_{t \to \infty}  \hat{\Phi}(x_1 ... x_t)  / t \,= \inf_{i=1, 2,  ... } 
  \lim_{t \to \infty}  | \varphi_i(x_1 ... x_t) | / t
  \end{equation}
   \end{theorem}
   A proof is given in Appendix 1, but here we note that Claim 1 is a particular case of this theorem. 

\textbf{Comment 3.}  If the sequence $x_1 x_2 ... $ is generated by a  stationary  source 
and all $\varphi_i$ are universal codes, 
the 
property iii)  is valid with probability 1 (See, for example, \cite{Cover:06}).
Hence, this theorem (and the claim 1) are valid for this case.

In general, 
the property iii)  
shows that the sequence under consideration has some stability. 
In turn, it gives a possibility to estimate 
characteristics of the whole sequence $x_1 x_2 ... $ based on its initial part.


\section{The time-universal code for   stationary ergodic  sources }
In this  
 subsection we describe a time-universal code (and the corresponding predictor) for stationary sources. It is based
 on optimal universal codes for Markov chains, developed by Krichevsky \cite{krichevsky1968relation,Krichevsky:93} and the 
 twice-universal code \cite{BRyabko:84}.
 Denote by $M_i$, $i = 1, 2, ... $ the set of Markov chains with memory (connectivity) 
 $i$, and let $M_0$ be the set of Bernoulli sources. 
 For stationary ergodic 
$\mu$ and an integer $r$ we denote by
$h_r(\mu)$ the $r$-order entropy (per letter) and let $h_\infty(\mu)$ be the limit entropy;
see for definitions \cite{Cover:06}.  

 Krichevsky  \cite{krichevsky1968relation,Krichevsky:93} 
 described the codes $\psi_0, \psi_1, ... $ which are asymptotically  optimal for  
$M_0, M_1, ... $, correspondingly.  
 If 
the sequence  $x_1 x_2 ... x_n$ , $x_i \in A$,  is generated by a  source $\mu$
$\in M_i$, the following inequalities are valid almost surely (a.s.):
 \begin{equation}\label{uc-mar}
 h_i(\mu) \le |\psi_i(x_1 ... x_t) | / t \,  \le  h_i(\mu) + ((|A| - 1) |A|^i + C)/t ,
  \end{equation}
where $t$ grows. (Here   $C$ is
a constant.)
The length of a codeword of the  twice-universal code $\rho$ is defined as the following "mixture":
\begin{equation}\label{uc-tu1}
| \rho(x_1 ... x_t) | = - \log \, \sum_{i=0}^\infty  \omega_{i+1} \, 2^{- |\psi_i(x_1 ... x_t) | } \,
\end{equation}
(It is well-known in Information Theory \cite{Cover:06} that there exists a code with such codeword lengths, because 
$\sum_{x_1 ... x_t \in A^t}$ $ 2^{-| \rho(x_1 ... x_t) |} $ $ =1$.)
This code is called twice-universal because for any $M_i$, $i=0, 1, ... $, and $\mu \in M_i$
the equality (\ref{uc-mar}) is valid (with different $C$). 
Besides, for any stationary ergodic source $\mu$ a.s.
\begin{equation}\label{u-st-er}
\lim_{t \to \infty}   |\rho_i(x_1 ... x_t) | / t \, =  h_\infty(\mu)   .
\end{equation}

Let us estimate the time of calculations necessary  when using $\rho$.
First,  note that 
it suffices to sum a finite number of terms  in (\ref{uc-tu1}), because all the terms $ 2^{- |\psi_i(x_1 ... x_t) | }$ 
are equal for $i \ge t$.  On the other hand, the number of different terms grows, where
$t \to \infty $ and, hence, the encoder should calculate  
$2^{- |\psi_i(x_1 ... x_t) | }$ for growing number $i$'s. 
It is known \cite{BRyabko:84 } that the time spent  for encoding one letter  is close for different codes 
$\psi_i$. 
Hence,  the time spent  for encoding one letter by the code $\rho$ grows to infinity, when 
$t$ grows.  The described below time-universal code $\Psi^\delta$ has the same asymptotic
performance, but the time spent  for encoding one letter is a constant.

In order to describe the time-universal code $\Psi^\delta$  we give some definitions.
Let, as before, $v$ be an upper-bound of the time spent  for encoding one letter by
any $\psi_i$,  $x_1 ... x_t$ be the generated word, 
\begin{equation}\label{T-del}
T = t \, v,  \,  N(t) = \delta T / v =  \delta \, t  , $$
$$ m(t) = \lfloor \log \log N(t)   \rfloor,  \, s(t)  = \lfloor  N(t)/m(t) \rfloor \, .
\end{equation}
Denote by $\Psi^\delta$ the following method:

\textbf{Step 1} Calculate $m(t), s(t) $ and
$$| \psi_0 (x_1 ... x_{s(t)})|, |\psi_1 (x_1 ... x_{s(t)})|, ... , |\psi_{m(t)} (x_1 ... x_{s(t)})| \, .
$$

\textbf{Step 2} Find such a $j$ that 
$$ - \log   |\psi_j (x_1 ... x_{s(t)})| = \min_{i = 0, ... , m(t)} |\psi_i (x_1 ... x_{s(t)})| .
$$

\textbf{Step 3}
Calculate the codeword $\psi_j (x_1 ... x_{t}) $ and 
output  $$  \Psi^\delta (x_1 ... x_{t}) =  \,
< j >  \psi_j (x_1 ... x_{t})  \, ,$$
where $<j>$ is  the $\lceil - \log \omega_{j+1} 	\rceil $-bit 
codeword of $j$.

The decoding is obvious. 
\begin{theorem} Let $x_1 x_2 ... $ be a sequence generated by a stationary source
and the code 
$  \Psi^\delta $ be applied.  Then this code is time-universal, i.e. a.s.
\begin{equation}\label{Th2}
\lim_{t \to \infty }   | \Psi^\delta (x_1 ... x_{t}) | /t =  \inf_{i = 0,1, ... }    \lim_{t \to \infty }
 | \psi_i (x_1 ... x_{t}) | /t \, .
\end{equation}
In the case of prediction
$$
\lim_{t \to \infty }   (  \log \Psi^\delta (x_1 ... x_{t}) ) /t =   \sup_{i = 0,1, ... }  \lim_{t \to \infty }
 ( \log \psi_i (x_1 ... x_{t}) ) /t \, .
$$
\end{theorem}
A proof is given in Appendix 1.

\section{ Conclusion}
Here we note some possible generalisations. 
We consider mainly the case of off-line prediction and data compression, where 
the whole sequence $x_1 ... x_n$ can be investigated in order to find a suitable data-compressor or predictor.
There are situations where the forecast should be made step-by-step, i.e. $x_{i+1}$
should be predicted based on $x_1... x_t$,  $x_{i+2}$ should be predicted based on
$x_1... x_{i+1}$ and so on.  The suggested approach and methods can be naturally 
extended to this case, too, if we take into account the possibility to store results of calculations
done on previous steps.

Another generalization is connected with the need to know in advance the speed of computing the forecast (or data compression). In such a case the goal of time-universal method can be the same:
it should limit the time of calculation by $T (1+\delta)$,  where $T$ is (unknown beforehand)
the time of the optimal method (from a given set). In such a case
the speeds can  be evaluated during the calculation. 

\section{Appendix 1 } 
Proof of Theorem 1.
Define $\lambda_i = \lim_{t \to \infty}  |\varphi_i(x_1 ... x_t)  | / t $,  and let
\begin{equation}\label{la-opt}
\lambda_o = \min_{i} \lambda_i , \,
\lim_{t \to \infty}
 | \varphi_{o}(x_1 ... x_t) | / t \, = \, \lambda_o\, .
  \end{equation}
   Let 
$\epsilon$ be any positive number. 
Having taken into account that the set $F$ is finite, from these definitions we can see that 
there exists such $t_1$  that
\begin{equation}\label{pr-i-eps-all}
|  \, \, | \varphi_{i}(x_1 ... x_t) | / t -\lambda_i|  < \epsilon \, \, for \, \varphi_i \in F, \, t > t_1 \, .
  \end{equation} 
 Taking into account the property i), we can see that there exists  such a number 
   $t_2$ for which $\tau_i(t)$ is defined for all $\varphi_i$ and $t > t_2$,  and
   denote $t_3$ $ = \max \{t_1,  t_2\}$.  Take any $n> t_3$ and
   suppose that a data-compressor $\varphi_s$
was chosen, when $\hat{\Phi}$ was applied to $x_1 x_2 ... x_n$.     
   Hence, from the property ii) we can see that there exists $t_4 > t_3$, such that 
   \begin{equation}\label{t5}
(-\log \omega_s +  | \varphi_{s}(x_1 ... x_{t_4}) | )/ t_4 \le (-\log \omega_o +  | \varphi_{o}(x_1 ... x_{t_4}) | )/ t_4 \, .
  \end{equation}
  From (\ref{la-opt}) we obtain the following two inequalities
  $$ 
   (-\log \omega_s +  | \varphi_{s}(x_1 ... x_{t_4}) | )/ t_4
   \ge \lambda_s - \epsilon \, , 
  $$
$$
(-\log \omega_o +  | \varphi_{o}(x_1 ... x_{t_4}) | )/ t_4 \,  \le \lambda_0 + \epsilon \, .
 $$  
  Having taken into account (\ref{t5})  we can see from the two latest  inequalities that
  $\lambda_s - \epsilon < \lambda_o + \epsilon $ and, hence, 
  $\lambda_s   < \lambda_o + 2 \epsilon $.
  Taking into account, that, by definition  (\ref{la-opt}),  
 $\lambda_o   < \lambda_s  $, we obtain
  \begin{equation}\label{t8} 
  \lambda_o   \le \lambda_s  <  \lambda_o + 2 \epsilon \, .
   \end{equation}
  Since $n > t_1$, we can see from   
  (\ref{pr-i-eps-all})  that
  $$
  \lambda_s - \epsilon < 
  (-\log \omega_s +  | \varphi_{s}(x_1 ... x_{n}) | )/ n
   < \lambda_s + \epsilon \, .
  $$
   Taking into account that 
  $(-\log \omega_s +  | \varphi_{s}(x_1 ... x_{n}) | )/ n$ $= \hat{\Phi}(x_1 ... x_{n})/n$
  we obtain from  (\ref{t8}) that
  $$
  \lambda_o - \epsilon < 
  \hat{\Phi}(x_1 ... x_{n})/n < \lambda_o + 3 \epsilon  \, ,
  $$ and, hence,
  $$
 \lambda_o - \epsilon <    \lim_{n \to \infty}
  \hat{\Phi}(x_1 ... x_{n})/n < \lambda_o + 3 \epsilon  \, .
  $$
  It is true for any $\epsilon > 0$,  hence, $ \lim_{n \to \infty}
  \hat{\Phi}(x_1 ... x_{n})/n =\lambda_o$. The theorem is proven.

 Proof of Theorem 2.
  It is known in Information Theory  \cite{Cover:06 }
   that  $ h_r(\mu)$ $ \ge h_{r+1}(\mu)$ $\ge h_\infty(\mu)$ for any $r$ and (by definition) 
   $\lim_{r \to \infty} h_r(\mu)$ $= h_\infty(\mu)$.
  Let   $\epsilon > 0 $ and $r$ be  such an  integer that $h_r - h_\infty$  $ < \epsilon$.
  From  (\ref{T-del}) we can see that there exists such $t_1$ 
  that $m(t) \ge r$ if $t \ge t_1$.  Taking into account (\ref{uc-mar}) and (\ref{T-del}), we
  can see that there exists $t_2$ for which a.s. 
  $| |\psi_r(x_1 ... x_t)| /t  - h_r(\mu) |$ $<\epsilon$ if $t > t_2$. From 
  the description of $\Psi^\delta$ (the step 3) we can see that there 
  exists such $t_3 > \max \{t_1, t_2\}$ for which a.s. 
  $$
  |  |\psi_r(x_1 ... x_t)| /t  - h_\infty(\mu) | \le  | |\psi_r(x_1 ... x_t)| /t  - h_r(\mu) |
  $$ $$ +
    ( h_r(\mu)  - h_\infty(\mu) ) < 2 \epsilon  \, ,
    $$
  if $t > t_3$. 
  By definition, 
$$  | \Psi^\delta(x_1 ... x_t)| /t \le  ( |\psi_r(x_1 ... x_t)|  - \log \omega_{r+1} ) /t .
$$
Having taken into account that $\epsilon$ 
is an arbitrary number and 
two latest inequalities as well as the fact that a.s.
$\inf_{i=0,1, ...} \lim_{t \to \infty }  |\psi_r(x_1 ... x_t)| /t $ $ = h_\infty (\mu)$,
 we obtain    (\ref{Th2}).
   The theorem is proven.

\section{ Appendix 2:  The Laplace approach to prediction  } 
Let there be  a source with unknown statistics which generates
sequences $x_1x_2\cdots$ of letters from some finite  alphabet
$A$.  Let the source generate a message $x_1\ldots
x_{t-1}x_t$, $\, x_i\in A$, and the following letter
$x_{t+1}$ needs to be predicted.
 This problem can be traced back to Laplace   who considered
the problem of estimation of the
 probability that the sun will
rise tomorrow, given that it has risen every day since Creation \cite{feller1958introduction}.
In our notation the alphabet $A$ contains two letters, $ 0 \; $ (``the
\:sun \:rises'')  and $1 \; $ (``the \:sun\: does\: not\: rise'' ); 
$t$ is the number of days since Creation, $x_1\ldots x_{t-1}x_t =
00 \ldots 0 .$

Laplace suggested the following predictor:
\begin{equation}\label{RAM:L}
L_0(a|x_1\cdots x_t) = (\nu_{x_1\cdots x_t}(a) +1 )/ (t+ |A | ),
\end{equation}   where $\nu_{x_1\cdots x_t}(a)$
denotes the count of letter $a$ occurring in the word $x_1\ldots
x_{t-1}x_t.$  For example, 
 if $ A= \{0, 1 \},$ $ \: x_1 ... x_5 $ $ = 01010$, 
then the Laplace prediction is as follows: $L_0(x_{6}=0| x_1 ...
x_5 =01010) $ $ = (3+1)/ (5+2) $ $ = 4/7, $ $ L_0(x_{6}=1 |x_1 ... x_5 = $  $ (2+1)/ (5+2) = 3/7.$ In other words, $3/7$ and  $4/7 $
are estimations of the unknown probabilities $P(x_{t+1} = 0|x_1
\ldots x_t = 01010 )$ and $P(x_{t+1} $ $ = $ $1 |x_1 \ldots x_t$ $ = $ $ 01010).$ 
(In what follows we will use the shorter notation: $P( 0| 01010
)$ and $P( 1 | 01010 ) ).$
We can see that Laplace  considered  prediction as a set of
estimations of unknown (conditional) probabilities, 
 because they contain all information about
 the
 future  behaviour of any stochastic process.  
In general,  we call as a predictor $\pi$ any conditional probabilities $\pi(x_{i+1} =a| $ $
x_1 = a_1,  ... , x_n = a_n)$  defined for all integers $n$, $a \in A$,
 $a_1, ... ,  a_n$ 
$\in A^n$. 

Proximity of the theory of universal coding and prediction,  as well as 
asymptotically optimal methods of prediction in a framework of the Laplace approach were found
for the cases of a finite alphabet and continues one in \cite{BRyabko:88} and \cite{BRyabko:09}, correspondingly.

\section*{Acknowledgment}
This work was supported by Russian Foundation for Basic Research (grant 15-07-01851).


\end{document}